\documentclass[aps,prl,twocolumn,superscriptaddress]{revtex4}
\usepackage{latexsym}
\usepackage{graphicx}
\usepackage{epsfig}
\newcommand \be {\begin{equation}}
\newcommand \bea {\begin{eqnarray}}
\newcommand \ee {\end{equation}}
\newcommand \eea {\end{eqnarray}}
\newcommand \bed {\begin{displaymath}}
\newcommand \eed {\end{displaymath}}

\newcommand{\bit}{\begin{itemize}}
\newcommand{\eit}{\end{itemize}}

\begin{document}

\title{ Tumor Gompertzian growth by cellular energetic balance}
\author{Paolo Castorina$^{2,}$}
\author{Dario Zappal\`a}
\affiliation{INFN, Sezione di Catania,
\quad and \quad
$^{2}$Dept. of Physics, University of Catania,\\
Via S. Sofia 64, I-95123, Catania, Italy
}
\date{\today}
\begin{abstract}
A macroscopic model of the tumor Gompertzian growth is proposed.
The new approach is based on the energetic balance among the different cell activities, 
described by methods of statistical mechanics 
and related to the growth inhibitor factors. The model is successfully applied 
to the multicellular tumor spheroid data.
\end{abstract}


\maketitle

A microscopic model of tumor growth in vivo is still an open problem. It requires a detailed description of cellular
interactions and a control on  the large variety of in situ conditions related to the distributions of nutrient, of oxygen,
of growth inhibitors, of blood vessels and capillarity and to the  mechanical effects due to tissue elasticity and heterogeneity.
On the other hand, in spite of the previous large set of potential parameters, tumors have a peculiar growth pattern
that is generally described by a Gompertzian curve \cite{gompertz}, often considered as a pure phenomenological fit to the data.
More precisely there is an initial  exponential growth
(until 1-3 mm in diameter) followed by the vascular Gompertzian phase\cite{libro}.
Then it seems reasonable to think that cancer growth follows a general pattern that one can hope to describe
by macroscopic variables, constrained by a set of environmental conditions crucial
to understand the fundamental features of the growth as, for example, the shape and the maximum possible size of the tumor and the
onset of the tissue invasion and metastasis.
Following this line of research, for example, the universal model proposed in \cite{west} has been recently applied to
cancer\cite{torino}.

In this paper we consider a macroscopic model of tumor growth that: i) gives an energetic basis to the Gompertzian law;
ii) clearly distinguishes among the general evolution patterns, 
which include feedback effects and  external constraints;
iii) can  give indications on the different tumor phases during its evolution.
The proposed macroscopic approach is not in competition with microscopic models \cite{varie}, but it 
is a complementary instrument for the description of the tumor growth.
The Gompertzian curve is solution of the  equation 
\be
\label{gomp}
 \frac{dN}{dt}= N \gamma \ln\left( \frac{N_{\infty}}{N} \right)
\ee
where $N(t)$ is the cell number at time $t$, $\gamma$ is a constant and $N_\infty$ is the theoretical saturation value for
$t \to \infty$. 

It is quite natural to identify the right hand side of  Eq. (\ref{gomp})  as the  number of proliferating cells at time 
$t$ and then to consider $f_p(N)= \gamma \ln\left( \frac{N_{\infty}}{N} \right)$ as the fraction of proliferating cells
and  $1-f_p(N)=f_{np}$  the fraction of non proliferating cells.
Here, we observe the difference with respect to an exponential growth that corresponds to a 
$N$ independent specific proliferation rate,
typical of a system of independent cells.
Instead, the logarithmic dependence observed 
in the Gompertzian  law, can be regarded as  a sort of  feedback mechanism, such that
the system,  at any time, rearranges the growth rate and the
fraction of  proliferating cells according to the logarithm of the 
total number of cells $N(t)$.

Living tumor cells consume an  amount of energy,  
under the form of nutrient and oxygen, which  is recovered from the 
environment (which will be addressed in the following as $B$), 
represented by a living body (in vivo) or an external solution (in vitro).
$B$, as already mentioned,  has also a mechanical
action   that can be responsible of compression, increase
of pressure and density  and  change of shape of the tumor
(addressed in the following as $A$).
Starting from  these observations, we intend to describe the tumor growth 
on an energetic basis.

Let us first consider the problem at fixed time and recall the first 
law of thermodynamics. A change of the energy content of a system 
has three possible origins, namely mechanical work performed on or by 
the system, variation of the number of elementary constituents (particles),
heat exchange with the environment. 
Accordingly, the internal energy of the  system is expressed as the sum of 
three terms, respectively $-PV$ (minus the product of pressure times volume),
$\mu N$ (where $\mu$, the chemical potential,  is the change in energy when 
the number of particles $N$ is increased by one unit), $S/\beta $ (heat content).

The basic ingredient in our model is an energy balance law for the tumor
in the same spirit of the first  law of thermodynamics.
However, it should be clear that  
the variables that we use to describe   our system
are not properly those introduced in thermodynamics but 
rather the rules and the relations among these variables  are 
the same as those of  thermodynamics.
In fact at some fixed time $t$ one can regard the available energy 
for the tumor, $U$, as the sum of three pieces, 
$U=\Omega + \mu N+E_M $,
where  $\Omega=-PV$ is associated to the mechanical work necessary 
to contrast the external pressure and to cause a  change of shape, 
$\mu N$ is proportional to the number of cells $N$ of $A$ and 
$\mu$ represents  the variation  in energy related to a change in  
$N$ (note, for the sake of clarity,  that the mechanical work related to the change 
in volume of $A$ due to the increasing number of cells is taken into 
account in $\Omega$)  and finally $E_M $, which represents the energy spent for 
metabolic activities.
Since the change in the number of cells in the tumor
is due to biological proliferation and by recalling 
the above definition of $\mu N$, 
it is natural to associate $\mu N$ to the energy 
that $A$ needs to generate one new cell, i. e. the energy necessary for reproduction,
that has  to be distinguished from the  energy associated to all the other metabolic 
processes which is addressed as $E_M$.

A set  of macroscopic variables suitable  to describe $A$ is:  
the volume of the whole tumor, $V$, the variable $\mu$ introduced above that 
we address as chemical potential, in analogy with the thermodynamical quantity, 
and $\beta$, analogous to  the  thermodynamical inverse temperature,
which is introduced,   together with its conjugate  variable $S$ 
(that corresponds to the thermodynamical entropy), by the relation $E_M=S/\beta$. 
The variables $V$, $\mu$ and  $\beta$ define the macroscopic state of $A$ at time $t$.

Statistical mechanics provides a framework that explains the basic laws 
of thermodynamics starting from a microscopic description of the 
thermodynamical system. 
Therefore, 
having in mind the energetic aspect of the problem,
we shall introduce a minimal microscopic  model of the cell,
where the biological aspects are reduced to the indispensable,
which however catches the essential ingredients 
that provide a sound macroscopic description and
a good comparison with the quantitative data.  
In an extremely simplified picture, 
we assume that the microscopic states of each cell
can be labelled by an integer index $l$ starting from a 
ground state level with  energy   $\epsilon$ so that each 
cell has an energy spectrum of the form 
$\epsilon_l = \epsilon  + l \delta$, where $\delta$ 
is the minimum positive energy gap between two  states. 
It should be clear that here  the energy of a cell indicates the energy 
that the cell is consuming per unit time, so that it is obvious that 
the minimum energy in the spectrum must be positive $\epsilon>0$.

According to  the basic rules of statistical mechanics,
once we have the energy spectrum of a single cell
we can determine the grand partition function,
${\cal Z}(\beta,V,\mu)=\Pi_{l=0}^\infty {\rm exp}\left( e^{-\beta(\epsilon_l-\mu)} \right )$
and the  grand potential,  corresponding to  the quantity $\Omega$
introduced above:
\be\label{omega}
\Omega(\beta,V,\mu)=-\frac{1}{\beta} \ln{\cal Z}= -\frac{W}{\beta}  e^{-\beta(\epsilon-\mu)}
\ee
where $W=1/(1-e^{-\beta\delta})$.
From the derivatives of $\Omega(\beta,V,\mu)$ we respectively get 
$N$, for constant $V$ and $\beta$, and $E_M$, for constant $V$ and $\mu$: 
\be\label{enne}
N=-\left (\frac{\partial \Omega}{\partial \mu}\right )_{V,\beta}= -\beta \Omega
\ee
\be\label{eemme}
E_M=\left (\beta \frac{\partial \Omega}{\partial \beta}\right)_{V,\mu}=\frac{N}{\beta}
\left ( 1+C+\ln\left( \frac{W}{N} \right ) \right )
\ee
where $C=W\beta\delta~{\rm exp}(-\beta\delta)$.
From  Eqs. (\ref{omega}) and (\ref{enne}) it is straightforward to express $\mu$ in terms of $N$ :
$\mu=\epsilon+(1/ \beta) \ln(N/W)$ and also to derive
the average energy per cell:  $U/N=(E_M +\mu N + \Omega)/N=\epsilon +C/ \beta$,
which shows that $U/N$ depends on one macroscopic variable only, the inverse temperature $\beta$,
as it is the case for an ideal gas in thermodynamics.

At this point we observe that there are two natural constraints in our model:
in fact it is clear that an increase in the number of cells $N$ requires a positive energy supply which 
has a direct consequence on the sign of 
the chemical potential: $\mu>0$.  Moreover $E_M>0$ because
the cells also require a positive energy supply for their metabolic
activity. In addition, in order to compensate any  mechanical stress on the tumor we need to require
a stronger constraint: $E_M+\Omega>0$. 
According to the relations derived above, the constraint $\mu > 0$ implies a minimum  number of cells 
$N<N_m= W~ {\rm exp}(-\beta \epsilon)$ and $(E_M+\Omega)/N=({1}/{\beta})\left ( C +
 \ln ( {W}/{N}) \right )>0$ 
provides a maximum:  $N< N_{\infty} =N_m {\rm exp} (C +\beta \epsilon)$.
Then, by replacing  $N_m$ and $N_\infty$ in the previous relations we get 
$\mu=(1/ \beta) \ln(N/N_m)$
and $(E_M+\Omega)/N =(1/ \beta)  \ln (N_{\infty}/N)$. Provided one identifies the 
maximum number of cells here obtained with the asymptotic value of $N$ appearing 
in Eq. (\ref{gomp}),  the latter equation for $(E_M+\Omega)/N $ is remarkably similar to the
the fraction of proliferating cells at time $t$, $f_p(t)$, introduced after Eq. (\ref{gomp})
and this will be our  starting point  to derive the Gompertzian law.

Now we assume that $A$, due to cellular biological duplication and possible changes 
induced by $B$, can modify its macroscopic state and therefore also 
the macroscopic variables which define the state itself
and determine the grand partition function {\cal Z}.
We assume that these changes occur on a time scale that is much larger than 
the typical  time during which the macroscopic variables reach the equilibrium and
become constant, 
so that we can think of the evolution of $A$ as a slow change from one
macroscopic state, where the statistical framework is well defined, to another one,
where the same framework, but with different values of the macroscopic variables, holds.

Recalling that at any time fixed $t$ one can write the identity $1= (f_p  + f_{np}) $ 
and the energy  sum rule is 
$U = ( {C/  \beta} + \epsilon) (f_p  + f_{np}) N = E_M+\Omega + \mu N$
and, according to the observed similarity between the expressions of $f_p$ and $(E_M+\Omega )/N$,
it looks natural to identify $E_M+\Omega=( {C/  \beta} + \epsilon) f_p N$ and, correspondingly 
$\mu=( {C/  \beta} + \epsilon) f_{np}$
This basic assumption is supported by a biological counterpart. In fact in \cite{suterland}
the feedback effect of Eq. (\ref{gomp}) is described  by introducing some 
biochemical effects due to growth inhibitor factors, released by the dead cells,
 which inhibit the proliferation and eventually are responsible 
of the saturation of the tumor size, so that the concentration of
these inhibitor factors should be  proportional to $f_{np}$.
In our energetic picture the concentration of 
inhibitor factors should act enhancing the energy necessary for increasing,
through cellular duplication, the number of cells in the tumor 
and therefore enhancing $\mu$. Then, the presence of growth 
inhibitor factors leads to a qualitative proportionality between 
$f_{np}$ and $\mu$ that supports our  assumption.

With the statement
$E_M+\Omega=( {C/  \beta} + \epsilon) f_p N$
it is straightforward to derive the Gompertz  equation.
In fact, by definition,  $\Delta N$  in an interval $\Delta t$
is proportional to the number of proliferating cells and then
$\Delta N=  c_1 \Delta t~f_p(t) N =c_2 \Delta t~(E_M+\Omega)$
with  $c_1$, $c_2$  constants. 
When $ \Delta t \to 0$ one recovers 
Eq. (\ref{gomp}) with $\gamma=c_2/\beta$.
Remarkably, 
this model explains the  linear correlation between the 
two standard parameters used to fit the data on tumor growth 
by the Gompertzian curve \cite{libro,param}.

A comment is in order. 
$N_m$ is related to $\mu=0$. In our set of relations  $N<N_m$ 
corresponds to $\mu<0$ which is not physically allowed. This indicates that our model 
and the feedback mechanism embedded in it, start being valid and hold only when $N \geq N_m$.
This again fits with the experimental observations that the onset of the Gompertzian growth 
does not occur at the beginning of the tumor development but only after an exponential phase growth \cite{libro}.
It is reasonable to think that  when $N<N_m$, due to the abundance of energy provided 
by $B$, the cells of $A$
duplicate as a system of  independent cells  with no further energetic effort for
the duplication due to the presence of the other cells,
i.e. with $\mu=0$ (and not $\mu<0$), which indicates that 
the feedback effect should not take place  and,  as noticed after Eq. (\ref{gomp}), this implies an exponential 
growth. Therefore, even if our model does not work for $N<N_m$, one reasonably  expects a continuous crossover  
between exponential and Gompertzian growth at $N=N_m$.

To analyze some  phenomenological implications of our  model we shall  consider 
in the following the simple case of the 
multicellular tumor spheroids (MTS), which follow the Gompertzian growth law as 
tumors in vivo, despite they  have an avascular growth.
We shall first consider experiments without external stress and, later,  
the effects due to a change of the  external pressure.
 
\vspace{0.5 pt}
 i) ENERGETIC GROWTH
\par\noindent
Here we shall refer to experiments where the variation of the external supply of nutrient and/or oxygen 
is studied  \cite{suterland,hel} and we shall rely on a minimal description of the MTS
where \cite{libro}:
a) the thickness, $k$, of the layer where
the nutrient and oxygen is delivered (the crust) is independent on the  spheroid radius $R$
($k$ as a function of $R$ is constant within $10\%$);
b) the cellular density is constant;
and essentially independent of the nutrient concentration;
c) the difference in  nutrient and oxygen  supply for  the cells in the inner and outer 
layers of the crust is negligible in first approximation,
d) the cells 
at  distances $d <R-k$ if $R>k$ from the center of the spheroid are dead and only the cells
of the crust are alive. 

The data, for various concentrations of glucose and oxygen
show that  \cite{suterland}: 
1) after an initial exponential growth (3-4 days) where the doubling time is independent 
of  the oxygen and glucose concentration,  the radius $R$ of MTS
grows according to a Gompertzian law  as the one in 
Eq. (\ref{gomp}), where  $N$ is replaced by $R$ and $ N_{\infty}$ by  $R_{\infty}$ ($R_{\infty}$ is the 
maximum radius  that the MTS reaches at saturation at very large $t$);
2) there is an approximate linear correlation between $k$  and  the logarithm of the glucose concentration, 
$G_c$, at fixed oxygen quantity;
3) there is a correlation between the radius at the onset of necrosis, $R=k$  and  at the growth saturation,
 $R=R_{\infty}$.

It is reasonable to think that at the early stages of the growth, when $R<k$,
all  cells receive the nutrient and oxygen supply while, when  $R>k$, due to a more difficult diffusion of the nutrient,
there is a fraction of non-proliferating cells and the feedback effect starts. According to our model,
one observes the Gompertzian phase for $t  \geq t^*$, with $t^*$  defined by the condition $R(t^*)=k$, when  the feedback begins.
Since  there is a simple geometrical relation between $N$ and $R$ due to point b),  it is easy to 
check that, for $R >>k$, our model predicts a Gompertzian growth for the MTS radius up to an asymptotic
value $R_{\infty}$, which is determined by $N_\infty$.
In our model, the number of cells at time $t^*$ is fixed by the condition $\mu =0$, i.e. $N(t^*)=N_m$ and,
as seen before, $N_m= N_{\infty} {\rm exp}(-\epsilon \beta -C)$. Putting all together these informations, 
we  find $k=R(t^*) \simeq N_{\infty}^{1/3}$. 

Since the experimental data at fixed oxygen quantity, are  almost consistent with  a  linear relation 
$N_{\infty}= a (G_c- G_c^0) + N_{\infty}^0$  where $a$,  $G_c^0$ and $N_{\infty}^0$ 
are ad hoc constants  (see Figure captions for details),
\cite{suterland,hel},
we are able to relate the MTS radius  at the onset on necrosis, $k$, to $G_c$ : 
\be\label{power}
k(G_c)=\alpha \left ( N_{\infty}^{1/3}(G_c)  - {N_{\infty}^{0}}^{1/3}\right )+ k_0
\ee
where $\alpha$ and $k_0$ are constants depending on the supplied oxygen. 
From Eq. (\ref{power}) one obtains the 
correlation among $N_{\infty},~G_c$ and $k$.
\begin{figure} 
\epsfig{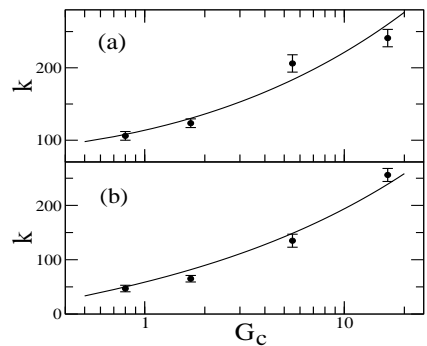} 
\caption{Thickness ($\mu m$) vs. glucose concentration ($mM$).
Figure (a) is for an oxygen concentration of $0.28~mM$ and obtained
for $N_{\infty}^{0}=2.1~10^5$, $a=1.8~10^5$, $G_c^0=0.8$.
Figure (b) is for an oxygen concentration of $0.07~mM$ and obtained
for $N_{\infty}^{0}=7.6~10^3$, $a=1.03~10^4$, $G_c^0=0.8$.
 } 
 \end{figure} 
In Fig.1 and  Fig. 2 the previous behaviors are compared with data 
without optimization of the parameters.
Despite  the oversimplified approach, there is a good agreement with the data.
Moreover the proposed model predicts the Gompertzian behavior on the basis of a feedback effect and a constant supply of nutrient for cell
and this explains why, under these conditions, the MTS and  the in situ tumors follow the same general growth  law. 

\begin{figure} 
 \epsfig{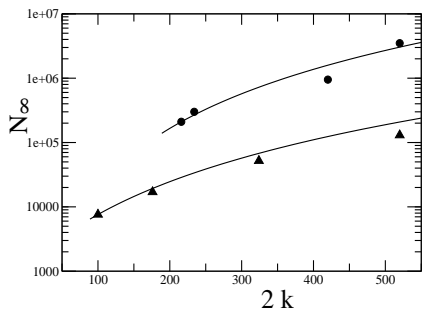} 
\caption{Spheroid saturation cells number vs. diameter ($\mu m$) at which necrosis first develops.
Circles  refer to culture in $0.28~mM$ of oxygen and the relative curve is obtained 
for $\alpha=1.78$ and $2 k_0=216 ~\mu m$.
Triangles  refer to culture in  $0.07~mM$  of oxygen and  in this case  $\alpha=5.3$, $2 k_0=100 ~\mu m$. } 
 \end{figure} 

\vspace{0.5 pt}

 ii) BIOMECHANICAL EFFECTS
\par\noindent 
The  heterogeneous elastic characteristics of the environment  give  the strongest external constraints\cite{brain}.
Again the MTS are useful to study how the mechanical features of the  environment can modify the growth pattern.
When  the MTS are under a solid stress, obtained for example  by a gel, the experimental data indicate that \cite{hel}:
1) an increase of    the external gel concentration $C_g$  inhibits the growth of MTS reducing
the size of $R_{\infty}$ with respect to the case with  $C_g=0$
(as before   $R_{\infty}$ is the 
maximum radius  that the MTS reaches at saturation);
2) the  cellular density at saturation  $\rho_{\infty}$ increases with  $C_g$.

In our model  the mechanical energy is included in the energetic balance by the term $\Omega$,
and therefore in this problem $\Omega$ depends on $C_g$. In particular, for growths in absence of gel, $C_g=0$, 
we neglect its value,  $\Omega\simeq 0$,  because of the very low pressure $P\simeq 0$, if compared to
the case where $C_g\neq 0$.
Also $N_{\infty}$ depends on $\Omega$ and, by 
recalling that  $N_{\infty}(\Omega)$ is obtained from the constraint $E_M+\Omega>0$
and therefore when $\Omega=0$ one can get $N_{\infty}(0)$ directly from $E_M>0$,
it is straightforward to derive 
$N_{\infty}(\Omega)= N_{\infty}(0)/e$.
For a growth with constant density this reduction should also imply
a decrease of the maximum size of the spheroids, i.e. $ R_{\infty}(\Omega)<R_{\infty}(0) $.
For a quantitative approach one has to take into account 
the correlation between the density $\rho$ and the dynamical growth
and, since $\Omega =  - N/\beta = - PV$, the  pressure is $P(t)= \rho(t)/\beta$. 
So, for $\rho_{\infty}$,  one has 
$ [\rho_{\infty}(\Omega)- \rho_{\infty}(0)] =\beta P_e$
where  $P_e$ is the external pressure due to  the gel $C_{g}$, and
$\rho_{\infty}{(0)}$ is the density for $P_e=0$, i.e. for $C_{g}=0$.
Then $\rho_{\infty}$ increases with  $C_{g}$ according to  point 2).
A linear fit to the data gives $\rho_{\infty}(\Omega)- \rho_{\infty}(0)=0.71 C_g$ \cite{hel}, and 
since  $\rho\sim N/R^3$, one still needs $N_{\infty}(\Omega)/N_{\infty}(0)$
to obtain the dependence of the MTS radius  at saturation $R_{\infty}(\Omega)$ on $C_g$.
The relation $N_{\infty}(\Omega)= N_{\infty}(0)/e$, derived above is discontinuous in the limit 
$\Omega\to 0$, which is an artifact of our  oversimplified microscopic model.
However,  for intermediate values of the gel concentration
we do not expect large deviations from this result and
we use it, in first approximation, to  determine $R_{\infty}(\Omega)$  in terms of $C_g$.
According to this procedure we have no adjustable parameters left  and 
the comparison with the data is reported in Table I for $C_g$ in the range $0.3 - 0.8~\%$.

\begin{table} [h] \centering{
\begin{tabular}{|c|c|c|}  \hline
\hline
$C_g$ (percent)& $2R_{\infty}(\Omega)$ $[\mu m]$  exper.& $2R_{\infty}(\Omega)$ $[\mu m]$ fit  \\
\hline
0.3  &  450  &  452 \\
0.5  &  414  &   429\\
0.7  &  370  &   404\\
0.8   &  363  &  394\\
\hline
\end{tabular}
\caption{\rm
Comparison with the experimental data as discussed in the text. The experimental error is about
$\pm 10 \%$ (see Fig. 1a in \cite{hel}).
}
\label{tab:1}
}
\end{table}

In conclusion, in the presented model  the Gompertzian growth  is essentially due to an average  energetic balance
which represents a macroscopic description of a
complex underlying dynamics. The comparison with the MTS 
data suggests that it is a good starting point for a macroscopic analysis of tumors in vivo
and indeed the Gompertz equation has interesting properties that can explain why this 
approximation works
in  realistic conditions \cite{noi}.

\begin{acknowledgments}
We thank  V. Albanese for discussions. 
We also thank Mrs. K. Corsino for the help in finding many quoted papers.
\end{acknowledgments}


\begin{thebibliography}{18}
\bibitem{gompertz} B. Gompertz, Phyl. Trans. R. Soc., {\bf 115} (1825) 513.
\bibitem{libro} G.G. Steel, ``Growth Kinetic of tumors'', Oxford Clarendon Press, 1977;
Cell tissue  Kinet.,  {\bf 13} (1980) 451;
T.E. Weldon, ``Mathematical models in cancer research'', Adam Hilger Publisher, 1988
and refs. therein.
\bibitem{west} G.B. West {\it et al.}, Nature {\bf 413} (2001) 628.
\bibitem{torino} C. Guiot {\it et al.}, J. Theor. Biol. {\bf 25} (2003) 147.
\bibitem{varie} M. Marusic  {\it et al.}, Cell Prolif. {\bf 27} (1994) 73; 
Z. Bajzer  {\it et al.}, `` Mathematical modelling of tumor growth kinetics'' in:
Survey of model for tumor-immune system dynamics, (J.A. Adams and N. Bellomo eds.) Birkhauser 1997;
A. Bru  {\it et al.}, Phys. Rev. Lett.  {\bf 81} (1998) 4008;
Z. Bajzer, Growth Dev. Aging, {\bf 63} (1999) 3;
N. Bellomo {\it et al.},  ``Mathematical topics on the modelling complex multicellular
systems and tumor immune cells competition'',  Preprint: Politecnico di Torino, 2004.
\bibitem{suterland} J. P. Freyer, R.M. Sutherland, Cancer Research,  {\bf 46} (1986) 3504.
\bibitem{param} L. Norton {\it et al.}, Nature {\bf 264} (1976) 542;  L. Norton Cancer Research,  {\bf 48} (1988) 7067.
\bibitem{hel} G. Helmlinger  {\it et al.},Nature Biotechnology,{\bf 15}(1997)778.
\bibitem{brain} G. Eaves, J. Path.,  {\bf 109} (1973) 233; R. Wasserman  {\it et al.}, Math. Bio.  {\bf 136} (1996) 111 and refs. therein.
\bibitem{noi} P. Castorina and D. Zappal\`a, in preparation.
\end{thebibliography}
\end{document}